\documentclass[preprint,aps,prl,unsortedaddress,superscriptaddress,showpacs]{revtex4}

\usepackage{amssymb}
\usepackage{graphicx}
\usepackage{dcolumn}
\usepackage{bm}
\usepackage{color}
\usepackage{ulem}

\begin{document}
\title{Terahertz response of carbon nanotube transistors}
\author{Diego Kienle}
\author{Fran\c{c}ois L\'{e}onard}
\date{\today }
\affiliation{Sandia National Laboratories, Livermore, CA 94550, USA}

\begin{abstract}
We present an approach for time-dependent quantum transport based on a
self-consistent non-equilibrium Green function formalism. The technique is
applied to a ballistic carbon nanotube transistor in the presence of a time
harmonic signal at the gate. In the ON state the dynamic conductance
exhibits plasmonic resonant peaks at terahertz frequencies. These vanish in
the OFF state, and the dynamic conductance displays smooth oscillations, a
signature of single particle quantum effects. We show that the nanotube
kinetic inductance plays an essential role in the high-frequency behavior.
\end{abstract}
\maketitle

Nanoelectronic devices using nanotubes and nanowires as their active
elements have been extensively studied for their DC properties~\cite{CNTDC}.
However, their high-frequency AC characteristics have received little
attention despite the obvious importance for many applications and the
breadth of intriguing scientific questions. Experimental work in carbon
nanotube (NT) and graphene field-effect transistors (FETs) have indicated
little performance degradation up to GHz frequencies~\cite{ExpRFCNT};
and recently, time-domain measurements in the terahertz (THz) regime have
been used to distinguish between plasmon and single-particle excitations at
low temperatures in NTs~\cite{McEuenNNT08}.

This progress in measuring the high-frequency properties of NTs poses
challenging questions for theory and modeling, in particular on how to
describe the carrier quantum dynamics under non-equilibrium conditions in
realistic device geometries, along with the self-consistent feedback between
the time-dependent charge and potential, which is essential to capture the
plasmonic excitations of the system~\cite{BohmPR49}. 
In addition, one open question is whether single-particle and plasmon excitations
can be distinguished in low-dimensional systems.

In this paper, we address these questions by proposing a formalism for AC
quantum transport making use of non-equilibrium Green functions 
(NEGF)~\cite{Haug1998}, and apply it to determine the high-frequency properties
of NTFETs. We find that the dynamic conductance exhibits both smooth
oscillations and divergent features in the THz regime, indicating the
coexistence of single-particle and collective excitations (plasmons). In
addition, by calculating the dynamic capacitance, we show that the nanotube
kinetic inductance plays a central role in determining the high-frequency
behavior, in contrast to conventional FETs.

We begin by describing our theoretical approach using the example of a NTFET
as illustrated in Fig. 1a. (NTs are ideal to study ballistic high-frequency
transport due to the large electron mean-free path for acoustic phonon
scattering, even at room temperature \cite{PurewalPRL07}.) The system is
divided into a device region which is connected to two semi-infinite leads
consisting of NTs embedded in source and drain metals. The salient feature
is the presence of a time-dependent potential at the gate terminal, 
$V_{g}+{\widetilde v}_{g}(t)$, and we are interested in the dynamic 
source-drain conductance $g(\omega )$.
This is a much different problem than the one treated previously  mostly within
a non-self-consistent approach where either a time-dependent 
potential was applied to the source and drain electrodes~\cite{TheoAC1} or 
simplified one- or two-level systems were used~\cite{TheoAC2}.
\begin{figure}[htbp]
\centering \includegraphics[width=8cm]{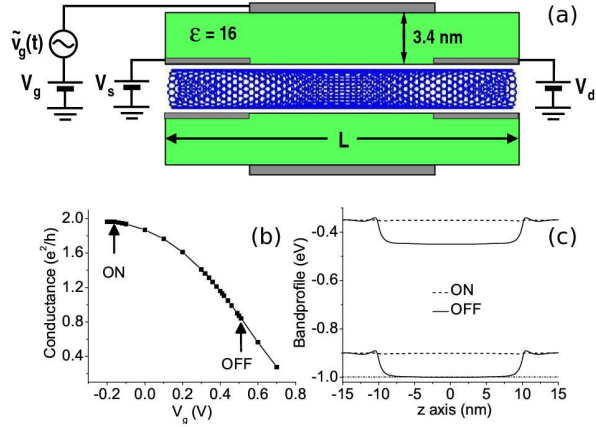} 
\vspace{-0.3cm}
\caption{(color online) (a) Cross section of the cylindrical NTFET geometry.
(b) DC transfer characteristics for a $L=30$ nm long NTFET. 
(c) Band diagram in the ON and OFF states, as marked by arrows in panel (b).}
\label{Fig1}
\end{figure}

In the absence of ${\widetilde v}_{g}(t)$, the system is described by the 
\textbf{r}etarded/\textbf{a}dvanced ($\gamma =r/a$) Green function 
\begin{equation}
G_{0}^{\gamma }(E)=\left[ \left( E\pm i\eta \right) \mathbf{I}
-H_{0}-U_{sc}-\Sigma _{c}^{\gamma }(E)\right] ^{-1}  \label{GreenDC}
\end{equation}
where $\eta $ is a positive infinitesimal constant. $H_{0}$ is the
time-independent Hamiltonian of the device region while $\Sigma _{c}^{\gamma}$ 
couples the device region and the semi-infinite leads; $U_{sc}$ is the
time-independent, spatially-dependent, self-consistent DC electrostatic
potential. This Green function forms the basis for calculations of the DC
transport properties of nanosystems. In the presence of ${\widetilde v}_{g}(t)$, 
a time- and spatially-dependent electrostatic potential $U(\mathbf{r},t)$ 
is generated in the device region. $U(\mathbf{r},t)$ must be determined based 
on the complex environment of the nanosystem, and must satisfy self-consistency 
between charge and electrostatic potential, as will be discussed further below. 
Because of the presence of this time-dependent potential, the relevant Green 
function is the two-time function
\begin{equation}
G^{\gamma }\left( t,t^{\prime }\right) = 
G_{0}^{\gamma } \left( t-t^{\prime} \right) 
+ \int dt_{1}G_{0}^{\gamma } \left(t-t_{1}\right) U(t_{1})
G^{\gamma}\left(t_{1},t^{\prime }\right) .
\end{equation}

We obtain the frequency-dependent \textit{particle} current at terminal 
$\alpha$ from the time derivative of the number operator as
\begin{widetext}
\begin{equation}\label{Current}
I_{\alpha }^{p}(\hbar \omega ) = \frac{e}{h}Tr\int dE\left\{
G^{<}(E^{+},E) \Sigma _{\alpha }^{a}(E) 
- \Sigma _{\alpha}^{r}(E^{+})G^{<}(E^{+},E)
+ G^{r}(E^{+},E)\Sigma _{\alpha }^{<}(E) 
- \Sigma_{\alpha }^{<}(E^{+})G^{a}(E^{+},E)\right\} ~,
\end{equation}
\end{widetext}
where $E^{+}\equiv E+\hbar \omega$. The AC Green functions 
$G^{\gamma,<}(E^{+},E)$ appearing in Eq.(\ref{Current}) are obtained from Dyson's 
equation $G^{\gamma }=G_{0}^{\gamma }+G_{0}^{\gamma }UG^{\gamma }$, whereas the
non-equilibrium particle density reads $G^{<}=(I+G^{r}U)G_{0}^{<}(I+UG^{a})$
following the standard procedure~\cite{Haug1998}. 
$G_{0}^{<}=G_{0}^{r}\Sigma_{c}^{<}G_{0}^{a}$ refers to the particle distribution 
of the unperturbed system with 
$\Sigma _{c}^{<}=-\sum_{\alpha }f_{\alpha }(\Sigma _{\alpha}^{r}-\Sigma _{\alpha }^{a})$, 
and $f_{\alpha }$ the Fermi function of terminal $\alpha $.

The above set of equations provides an approach to calculate the frequency-dependent 
quantum transport in the NEGF technique. These equations need to be augmented to include 
the coupling of the quantum transport equations with the electrostatics, as embodied 
by Poisson's equation:
\begin{equation}\label{poisson}
\nabla \cdot \left[ \epsilon (\mathbf{r})\nabla U(\mathbf{r},\omega )\right]
=-\rho (\mathbf{r},\omega ).  
\end{equation}
(At the frequencies of interest here, the time dependence of the full
Maxwell equations can be neglected.) This contribution is especially
important to capture the complex environments of nanoelectronic devices, and
plasmonic effects. Poisson's equation is complemented by appropriate boundary
conditions at the terminals; in particular at the gate terminal, a
time-dependent potential $\widetilde{v}_{g}(t)$ is applied and serves as the
external perturbation that generates $U(\mathbf{r},\omega )$.

A closed set of equations can be obtained by expressing the charge density
from the Green function:
\begin{equation}
\rho (\mathbf{r},\omega )=ie\pi ^{-1}\int dEG^{<}(E+\hbar \omega ,E).
\label{rho}
\end{equation}
Thus, the set of Eqs. $\left(\ref{GreenDC}\right) - \left( \ref{rho}\right) $
provides an approach to calculate self-consistently the frequency-dependent
current in the presence of a frequency-dependent external gate potential. We
note that our approach does not rely on approximations to the wide- or
narrow-band limits, but treats the spectral properties of the contacts explicitly.

To proceed further, we consider $U(\mathbf{r},\omega )$ to be a small
perturbation and expand the above equations to linear order in $U$. The
Green function of the perturbed system can be obtained from Dyson's equation
as $G^{\gamma }=G_{0}^{\gamma }+\frac{1}{2}G_{0}^{\gamma ,+}UG_{0}^{\gamma }$, 
while $G^{<}=G_{0}^{<}+\frac{1}{2}G_{0}^{<,+}UG_{0}^{a}
+\frac{1}{2}G_{0}^{r,+}UG_{0}^{<}$ (a $+$ superscript indicates a function evaluated
at $E+\hbar \omega $). In practice, one is often interested in the small-signal
response of the two-terminal conductance 
$g_{\alpha \beta }\left(\omega\right) = 
\left. dI_{\alpha }(\omega )/dV_{\beta }\right|_{V_{\beta }=0}$.
To obtain an expression for $g_{\alpha \beta }$, we consider a time-harmonic
gate potential ${\widetilde v}_{g}(t) = v_{0}\cos \left( \omega t\right)$ which
generates an electrostatic potential $U(\mathbf{r},\omega )\cos\left(\omega t\right)$ 
on the NT. Expanding to lowest order in $U$ and $V_{\beta }$ we obtain for the 
\textit{particle} conductance
\begin{widetext}
\begin{eqnarray}
g_{\alpha \beta }^{p}\left( \omega \right) &=& \frac{e^{2}}{2h}Tr\int dE
\left[ \left\{ G_{0}^{r,+} U \left( \omega \right) G_{0}^{r} \widetilde{\Sigma }_{\beta }^{<}
-\widetilde{\Sigma }_{\beta }^{<,+} G_{0}^{a,+} U \left(\omega\right) G_{0}^{a}\right\} 
\delta_{\alpha\beta }
+\widetilde{G}_{0,\beta}^{<,+} U \left(\omega\right) G_{0}^{a} \Sigma_{\alpha }^{a}\right.\nonumber \\
&&\left. + G_{0}^{r,+} U \left( \omega \right) \widetilde{G}_{0,\beta}^{<} \Sigma _{\alpha }^{a}
-\Sigma _{\alpha }^{r,+}\widetilde{G}_{0,\beta}^{<,+} U \left(\omega\right) G_{0}^{a}
-\Sigma _{\alpha }^{r,+}G_{0,\beta}^{r,+} U \left(\omega\right) \widetilde{G}_{0,\beta }^{<}\right] 
\end{eqnarray}
\end{widetext}
where $\widetilde{h}_{\beta }=\left. \partial h/\partial 
V_{\beta }\right|_{V_{\beta }=0}$ for a general function $h$.

In general, the particle conductance does not obey fundamental sum rules for
current conservation and gauge invariance, since the \textit{displacement
current} has been omitted. To include it, we adopt the scheme of Wang 
\textit{et.al.}~\cite{GuoPRL99}, where the final conductance is given by 
$g_{\alpha\beta} = g_{\alpha\beta}^{p} - \left( \sum_{\gamma =s,d}
g_{\alpha\gamma}^{p}/\sum_{\delta=s,d} g_{\delta }^{d}\right) g_{\beta }^{d}$,
and derive for the frequency-dependent displacement conductance 
\begin{eqnarray}
g_{\beta }^{d}(\omega ) &=& \frac{e^{2}{\omega }}{4\pi }\int dE
\left[ G_{0}^{r,+} \tilde{\Sigma}_{\beta }^{<,+} G_{0}^{a,+} U({\omega }) G_{0}^{a } 
\right. \nonumber \\
&&\left. + G_{0}^{r,+} U({\omega }) G_{0}^{r} \tilde{\Sigma}_{\beta}^{< }G_{0}^{a}\right]~.
\end{eqnarray}
Note that $g_{\alpha\beta}(\omega)$ measures the change in the conductance 
relative to the value at the DC operation point.

We now apply the AC theory outlined above to calculate the zero-bias AC
conductance at $T=300$ K for the NTFET of Fig. 1a. The first step is to
obtain the DC properties of the NTFET; details of the numerical procedures
are given in Ref.~\cite{LeonNT06}. The (17,0) NT of radius $R=0.66$ nm is
modeled using a $p_{z}$-tight-binding model with an overlap energy of 
$\gamma _{0}=2.5$ eV, giving a bandgap of $0.55$ eV. The Fermi levels $E_{F}$
of the source and drain metals are set 1 eV below the NT midgap before
self-consistency, which gives a p-type ohmic contact after self-consistency.
Fig. 1b shows the DC transfer characteristics of the NTFET with a channel
length $L=30$ nm, which consist of an ON-state where the bands are
essentially flat, and an OFF-state where a gate-controlled barrier blocks
the hole current (Fig. 1c).
\begin{figure}[htbp]
\centering\includegraphics[width=8cm]{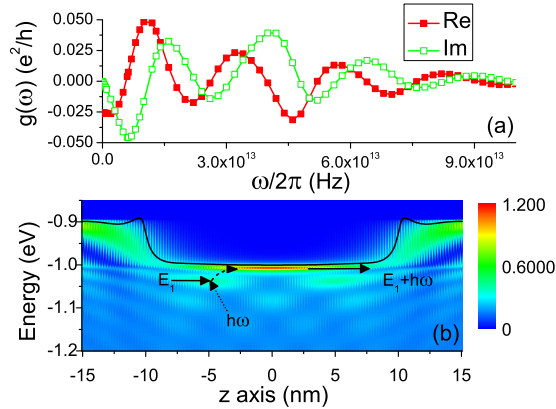} \vspace{-0.6cm}
\caption{(color online). AC Response in the OFF-state for a $L=30$ nm NTFET:
(a) Real/imaginary part of $g(\omega )$. (b) Color plot of the
density of states along the channel, showing the resonant photoexcitation of
carriers through energy and spatially oscillating quantum states. The
valence band edge is marked with the solid black line, and $E_{F}=-1$ eV.}
\label{Fig2}
\end{figure}
To explore the AC behavior of the NTFET, we first choose a DC operating
point, either in the ON or OFF state as marked by arrows in Fig.~\ref{Fig1}b, 
and apply an AC gate signal perturbation of frequency $\omega $ and magnitude 
$v_{0}=10$ meV. Figure~\ref{Fig2}a shows the real and imaginary parts of the AC
conductance in the OFF state which displays smooth oscillations as a function 
of frequency. ($g(\omega=0)$ is negative because the AC perturbation 
$v_0 \cos(\omega t)$ reduces to a positive DC voltage perturbation $\delta V_g = v_0$. 
According to the DC transfer characteristics of Fig.~\ref{Fig1}b an increase in 
DC gate bias leads to a reduced conductance.)
Surprisingly, for finite frequencies the AC conductance can take values \textit{larger}
than the DC conductance. The origin of this behavior and of the smooth oscillations 
can be understood from the spatially and energy dependent density of states (DOS) in 
the OFF state shown in Fig.~\ref{Fig2}b.
At a given position along the NT the DOS shows oscillations in energy with a 
characteristic frequency of about $25$ THz. The maxima in $g(\omega )$ arise 
from the photoexcitation of carriers between maxima of the DOS while the minima 
in $g(\omega )$ arise from the excitation between maxima and minima of the DOS. 
Thus, in the OFF state, smooth conductance oscillations are a signature of the 
single-particle excitation spectrum. The oscillatory character of $g(\omega )$ is 
preserved when the self-consistent feedback between charge and potential is disabled 
(not shown), confirming this nature of transport.
\begin{figure}[htbp]
\centering\includegraphics[width=8cm]{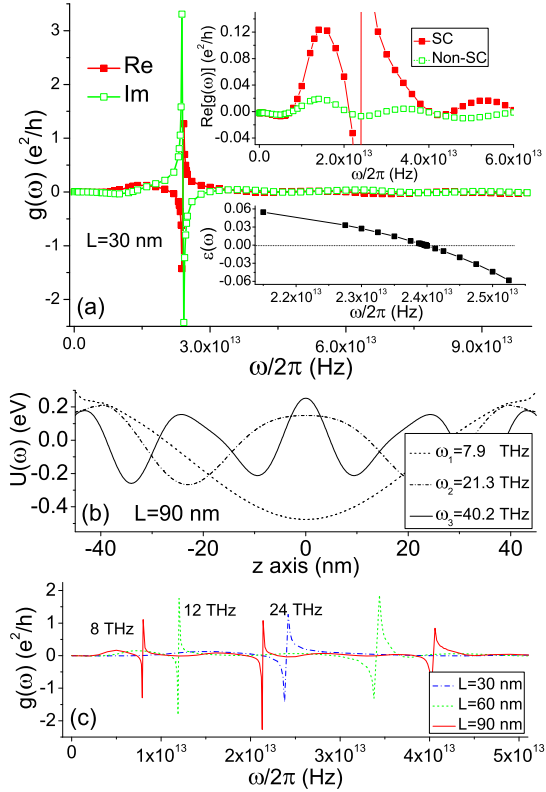} \vspace{-0.6cm}
\caption{(color online) Panel (a) shows the frequency-dependent conductance
for a NTFET with $L=30$ nm. Top inset compares the self-consistent and
non-self-consistent results. Bottom inset shows the dielectric function
versus frequency. Panel (b) shows the periodic potential near the plasmon
frequencies for $L=90$ nm. Panel (c) shows the response for three different
channel lengths. }
\label{Fig3}
\end{figure}

Figure 3a displays the real and imaginary parts of the dynamic conductance 
$g(\omega)$ for the same NTFET in the ON-state, which is also slightly 
negative at $\omega=0$ for the same reason as in the OFF-state. 
We observe that the response for frequencies less than about $4$ THz is 
constant, in agreement with the experiments in Ref.~\cite{ExpRFCNT}.
For larger frequencies we observe that the conductance exhibits a pronounced 
divergent response at $\omega/2\pi \approx 24$ THz with an underlying 
oscillatory behavior. Near the divergence, the self-consistent potential
along the NT reveals large amplitude oscillations, which change phase as the
divergence is crossed. These large amplitude oscillations are shown for a
device with $L=90$ nm channel length in Fig. 3b, including the higher-order
modes. The presence of these oscillatory modes suggests that plasmons are
responsible for the divergent behavior.
In order to ascertain that the response calculated in the ON-state can be 
attributed to \textit{collective} rather than single particle excitations, 
we compare in Fig.~\ref{Fig3}a (top inset) the self-consistent (SC) response
$g(\omega)$ with the non-self-consistent solution obtained from the first iteration. 
The divergence in the conductance disappears entirely while the smooth oscillations 
persist. Thus, the self-consistency between the charge and potential is essential 
to observe the divergence, a signature of a collective phenomenon. In the ON 
state we therefore have the coexistence of single-particle and plasmonic effects.

Furthermore, from the amplitude of the potential oscillations we can obtain
the dielectric function of the NTFET from $\epsilon (\omega )=U_{ext}\left(
\omega \right) /U\left( \omega \right) $. As shown in the bottom inset of 
Fig.~\ref{Fig3}a, $\epsilon (\omega )$ shows a clear zero crossing at the frequency 
where $g \left(\omega\right)$ displays divergent behavior, and is further
evidence for the excitation of plasmons. In the simplest model, $\varepsilon
\left( q,\omega \right) =1-\chi \left( q,\omega \right) K\left( q\right) $
where $\chi \left( q,\omega \right) $ is the response function and 
$K\left(q\right)$ is the electrostatic Green function for Poisson's equation. 
In the ON state, the NT is effectively like a quasi-one-dimensional electron
gas, where for small $q$, 
$\chi \left(q,\omega\right)\rightarrow 2q^{2}k_{F} / m\pi \omega ^{2}$, with 
$k_{F}$ the Fermi wavevector and $m$ the effective mass~\cite{Hu}. 
$K(q)$ is obtained by using the Green function for a NT in free space surrounded 
by a dielectric $\varepsilon_{ox}$, $K(q) = \frac{e^{2}}{\varepsilon_{ox}}I_{0}(qR)K_{0}(qR)$. 
(This is a good approximation to the full $K(q)$ when $q\gg R_{G}^{-1}$ where 
$R_{g}=4$ nm is the gate radius.) Near the threshold voltage we obtain the plasmon 
frequencies for the NTFET by quantizing the plasmon excitations to the channel length, 
i.e. $q_{n} = (\pi /L) \left( 2n+1 \right)$, giving the frequencies 
$\omega_{n} = \frac{v_{p}}{L} \left( 2n+1 \right) \sqrt{I_{0}\left( n\pi R/L \right) K_{0}I_{0} 
\left( n\pi R/L \right)}$ where $v_{p}$ is the plasmon group velocity. 
To test the scaling of the plasmon frequencies with channel length and mode number, 
we calculated the AC response for NTFETs up to channel lengths of $90$ nm; 
inspection of Fig.~\ref{Fig3}c indicates that the plasmon frequencies scale linearly 
with $L^{-1}$ and $n$, with a plasmon velocity of $v_{p}=2.7\times 10^{6}$ m/s, 
about four times larger than the Fermi velocity.

\begin{figure}[htbp]
\centering\includegraphics[width=8cm]{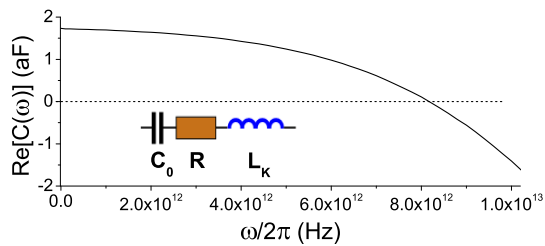} \vspace{-0.6cm}
\caption{(color online). Real part of the dynamic capacitance 
$C(\omega )$ for $L=30$ nm in the ON state. 
Inset: RLC-model for the dynamic capacitance $C(\omega)$.}\label{Fig4}
\end{figure}

Our modeling approach also allows the study of the fundamental processes
that control the AC properties of nanoelectronic devices. As an example, we
show in Fig.~\ref{Fig4} the real part of the total dynamic capacitance defined 
as $C(\omega )=Q(\omega )/v_{0}$ with $Q$ the total charge on the NT. 
At small frequencies $\mbox{Re}[C(\omega)]$ is positive implying a capacitive-like 
behavior, but becomes inductive at $\approx 8$ THz as marked by the sign change. 
This characteristics is fundamentally different from that of traditional FETs which 
show only capacitive behavior. 
The origin of this unconventional behavior is the nanotube kinetic inductance. 
Indeed, we can model the NTFET as a classical RLC circuit as shown in Fig.~\ref{Fig4} 
(inset), for which the dynamic capacitance 
$\mbox{Re}[C(\omega)] \varpropto 1-\omega^{2} L_{K} C_{0}$ where $L_{K}$ is the 
nanotube kinetic inductance and $C_{0}$ is the zero-frequency capacitance. 
This gives a transition to inductive behavior at $\omega = 1/\sqrt{L_{K}C_{0}}$, and 
from our numerical data for $C(\omega )$ we extract $L_{K}=0.2$ nH for the 30 nm device.
This value is consistent with that expected from simple arguments~\cite{book}.

In summary, we presented a new self-consistent approach for AC quantum
transport and applied it to determine the high-frequency response of NTFETs.
In the ON-state, the dynamic conductance shows divergent peaks, which are 
associated with the excitation of plasmons of the gated NTFET acting as resonant 
quantum cavity, whose mode spectrum can be tuned by varying the channel length. 
Our results suggest that low-dimensional systems with nanometer sized channels 
show potential for novel detectors and emitters of terahertz radiation.
The approach can be applied to a broad range of nanoelectronic systems; it
will be useful for the study of many time-dependent phenomena in
low-dimensionality systems including phonon and defect scattering, ultrafast
optical excitation, and entirely new operation modes of nanoelectronic
devices.

We are indebted to M. Vaidyanathan and H. Guo for fruitful discussions.
Sandia is a multiprogram laboratory operated by Sandia Corporation, a
Lockheed Martin Co., for the United States Department of Energy under
Contract No. DEAC01-94-AL85000.

\end{document}